\documentclass{optica-article}

\journal{opticajournal} 

\articletype{Research Article}

\newcommand{\figwidthPN}{11cm}  

\usepackage{lineno}

\begin{document}

\title{Dual-Wavelength Brillouin Lasers as compact Opto-Terahertz References for Low-Noise Microwave Synthesis}

\author{Scott C. Egbert,\authormark{*} James Greenberg, Brendan M. Heffernan, William F. McGrew, Antoine Rolland}

\address{Boulder Research Labs, IMRA, America Inc., 1551 S. Sunset Street, Longmont, 80501, CO, USA.\\}

\email{\authormark{*}segbert@IMRA.com} 


\begin{abstract}
Compact, ultra-low phase noise 10\,GHz signals are essential for modern radar, coherent communications, and time-frequency metrology, especially with rising demands  for additional spectral purity and portability. Optical frequency division (OFD) of ultra-stable optical references produce the lowest noise microwaves, but typically rely on ultra-low-expansion cavities, self-referenced frequency combs, pulse interleaving, and high-end photodetectors. In contrast, electro-optic frequency division (eOFD) offers a streamlined alternative in which an electro-optic (EO) comb is generated from a microwave source and stabilized to an optically carried terahertz (opto-terahertz) reference. eOFD has already demonstrated comparable phase noise to OFD at 10, 20, and 40\,GHz when using division ratios from references spanning over 1\,THz, requiring broad EO comb spectra to bridge between optical signals. Generating such broad spectra often demands complex techniques such as cascaded modulators, pulse compression, and nonlinear fiber. We demonstrate a compact, low-noise 300\,GHz opto-terahertz reference utilizing a dual-wavelength Brillouin laser within a total system volume of 20\,liters. The 300\,GHz phase noise of this system is transferred to a 10\,GHz dielectric resonant oscillator via eOFD using a simple architecture that could be miniaturized. The resulting microwave signal achieves phase noise levels of --130\,dBc/Hz at 1\,kHz, --150\,dBc/Hz at 10\,kHz, and --170\,dBc/Hz at 10\,MHz. This architecture drastically simplifies eOFD while maintaining state-of-the-art phase noise.
\end{abstract}

\section{Introduction}

The generation of ultra-low phase noise microwave signals has direct benefits to diverse fields: improving
resolution in radar~\cite{ghelfi2014fully}, spatial accuracy for navigation~\cite{OpticalClocksAtSea}, phase coherence in interferometric radio astronomy~\cite{Caldwell2023VLBI, Clivati2017VLBI}, capacity of coherent communications~\cite{kurner2016thz}, and fidelity in atomic and molecular spectroscopy~\cite{townes1975microwave,patterson2013enantiomer}, in addition to precision time and frequency metrology~\cite{rubiola2008phase}. 
Many state-of-the-art applications place stringent requirements on signal spectral purity, frequency stability, and phase coherence, further complicated by the desire to realize compact and field-deployable systems.

Among the techniques developed to meet these demands, OFD of an ultra-stable optical reference has established itself as the most effective method for synthesizing microwave signals with unmatched phase noise performance~\cite{millo2009ultra,fortier2011generation}. In a typical OFD system, a self-referenced optical frequency comb is phase-locked to a cavity-stabilized laser, enabling optical phase stability to be coherently transferred down to microwave frequencies. Pulse interleaving, high-speed photodetection, and narrowband filtering are then employed to extract the desired microwave tone~\cite{haboucha2011optical,quinlan2014optical,kalubovilage2020ultra}. This architecture has demonstrated record-low phase noise, approaching the thermal noise floor of room-temperature electronics~\cite{quinlan2013exploiting,xie2017photonic}, and remains a gold standard for low-noise microwave generation in laboratory environments. However, these systems are constrained by their size and reliance on sensitive, high-performance optical components, which limit their use in portable or integrated applications.

To overcome these limitations, several simplified OFD schemes have been proposed. Two-point OFD removes the need for self-referencing by locking two optical modes of a frequency comb to a pair of optical references~\cite{swann2011microwave,tetsumoto2021optically,liu2024low}, enabling division without detecting the carrier-envelope offset. Although the complexity is reduced, this technique still generally relies on broad comb spans enabled by lab-scale components, especially at high division ratios. 

Over the last decade, eOFD has emerged as an attractive alternative that eliminates the need for self-referencing and state-of-the-art photodetection~\cite{li2014electro}. In this approach, a microwave oscillator drives an EO comb generator with the resultant EO comb stabilized to an opto-terahertz reference using a phase-locked loop (PLL). Unlike Kerr-based or mode-locked combs, electro-optic combs are relatively insensitive to environmental changes and offer flexible repetition rates, governed by the characteristics of the microwave oscillator that drives the modulators. Additionally, eOFD enables the direct transfer of optical coherence to microwave frequencies without carrier-envelope offset detection, and has demonstrated low phase noise at 10\,GHz, 20\,GHz, and 40\,GHz~\cite{li2023small}.

Despite its advantages, eOFD faces a key limitation: the EO comb must span a sufficient bandwidth to cover the frequency gap between the optical reference and the microwave target frequency with sufficient optical signal-to-noise ratio (SNR) to preserve signal integrity. Demonstrations to date have typically used optical references spaced by more than 500\,GHz~\cite{kudelin2024tunable,he2024highly}, improving stability by increasing the division factor at the cost of needing EO combs with very broad spectra. Generating such broad combs requires a combination of multiple cascaded phase modulators, pulse compression stages, and/or propagation through highly nonlinear fiber~\cite{loh2024ultralow}, which reintroduces a high component count, reduces optical signal strength, and increases the overall Size, Weight, and Power (SWaP) requirements of the system.

In this work, we present a compact and streamlined eOFD architecture that circumvents this limitation by leveraging a low-noise opto-terahertz reference generated via stimulated Brillouin scattering in an 80-meter-long fiber cavity, as described previously by Heffernan et al.~\cite{heffernan_brillouin_2024}. The exceptionally low phase noise of the dual-wavelength Brillouin laser (DWBL) enables frequency division at a relatively small optical spacing of 300\,GHz. This moderate span can be bridged using a single EO phase modulator, which preserves high optical signal-to-noise ratio (SNR) by minimizing insertion loss and avoiding the additional components and RF drive requirements associated with cascaded modulators or nonlinear broadening stages. As a result, the beatnote can be detected cleanly, allowing stable phase locking of the microwave oscillator driving the EOM to the opto-terahertz reference. The synthesized 10\,GHz signal exhibits phase noise levels of --130\,dBc/Hz at 1\,kHz, --150\,dBc/Hz at 10\,kHz, and --170\,dBc/Hz at 10\,MHz offset. While these values are comparable to the best reported from traditional eOFD systems, further improvement at intermediate offsets (e.g., 100\,kHz to 1\,MHz) is currently limited by a servo bump associated with the loop bandwidth. We also characterize the phase noise of a similar DWBL operating at 3.33\,THz and evaluate its projected performance if divided down to 10\,GHz. These results highlight the scalability of low-noise opto-terahertz references for compact eOFD architectures. Potential paths toward integrating this 3.33\,THz reference into a deployable eOFD microwave synthesizer are discussed later in the paper.

\section{eOFD Phase Noise Suppression via Dual-Sideband Beat Detection}
\label{sec:equations}

Here we derive the phase noise power spectral density (PSD) \( S_{\phi}(f) \) of the output signal in an eOFD architecture, and highlight structural considerations for minimizing amplifier-induced noise. A schematic of the system used in these experiments is shown in Fig. \ref{fig:SchematicEOFD} \textbf{(a)}. This analysis emphasizes the benefits of using a single modulator and amplifier to enable full cancellation of common-mode phase noise within the phase-locked loop, similar to that of Li et al.~\cite{li2014electro}.

\subsection{System Architecture and Beatnote Formation}

Consider two continuous-wave (CW) optical signals \( \nu_{1} \) and \( \nu_{2} \), separated by a large optical spacing,
\begin{equation} \label{eq:nu}
\Delta \nu = \nu_{2} - \nu_{1}, \quad \text{with } \Delta \nu \gg 10~\mathrm{GHz}.
\end{equation}
These signals are simultaneously modulated by the same electro-optic phase modulator (EOM), driven at frequency \( f_{\mathrm{\upmu w}} \) by a microwave reference, for instance a dielectric resonator oscillator (DRO). This generates sidebands,
\begin{equation}
\nu_{1}^{(n)} = \nu_{1} + n f_{\mathrm{\upmu w}}, \qquad \nu_{2}^{(-m)} = \nu_{2} - m f_{\mathrm{\upmu w}}.
\end{equation}
Where these sidebands spectrally overlap, a beatnote is detected at,
\begin{equation} \label{eq:if}
f_{\mathrm{IF}} = \left| \Delta \nu - (m+n) f_{\mathrm{\upmu w}} \right|.
\end{equation}
This intermediate frequency is compared to a reference oscillator \( f_{\mathrm{ref}} \), and the resulting error signal is used to phase-lock the microwave reference such that,
\begin{equation}
f_{\mathrm{\upmu w}} = \frac{\Delta \nu \mp f_{\mathrm{ref}}}{m+n},
\end{equation}

\noindent where the sign of $f_\mathrm{ref}$ is opposite the sign of $(\Delta \nu - (m+n) f_{\mathrm{\upmu w}})$.

\subsection{Phase Noise Transfer Function}

To express the phase noise of the output signal in terms of the various contributing components, we evaluate how phase fluctuations propagate through this feedback system.

Treating the optical frequency difference \( \Delta \nu \) as a single term (e.g. DWBL) and assuming otherwise uncorrelated noise sources, the total phase noise PSD of the locked microwave signal amounts to,
\begin{equation}
S_{\phi,\mathrm{\upmu w}}(f) \propto \frac{1}{(m+n)^2} \left[
    S_{\phi,\Delta \nu}(f) + S_{\phi,\mathrm{ref}}(f) + S_{\phi,\mathrm{loop}}(f)
\right],
\label{eq:Sphi_DRO}
\end{equation}

\noindent where $S_{\phi, x}(f)$ is the one-sided phase noise PSD from noise source $x$. $\Delta \nu$ refers to the noise from the dual-wavelength source; $\mathrm{ref}$, the reference oscillator noise; and loop, residual servo noise (e.g., mixer, delay, detector).

It is important to note that the 10\,GHz signal used to drive the EOM is amplified by a high-power amplifier (HPA) in order to generate a sufficient number of higher-order sidebands to bridge between the optical lines with high SNR. Whether the noise of the HPA, $S_\mathrm{\phi,HPA}(f)$, is preserved or canceled in the DRO output depends on where the 10\,GHz signal is tapped. If it is tapped \textit{after} the amplifier, the tapped signal is the same that is servoed, and the phase noise is given by Eq. (\ref{eq:Sphi_DRO}). Conversely, if it is tapped \textit{before} amplification, the phase is given by $\phi_\mathrm{\upmu w}-\phi_\mathrm{HPA}$, and the phase noise is therefore,

\begin{equation}
S_{\phi,\mathrm{preamp}}(f) \propto \frac{1}{(m+n)^2} \left[
    S_{\phi,\Delta \nu}(f) + S_{\phi,\mathrm{ref}}(f) + S_{\phi,\mathrm{loop}}(f)
\right] + S_\mathrm{\phi, HPA}(f).
\end{equation}

\noindent We note that the phase noise from the HPA is not, as is the case for noise from other sources, reduced by the factor $(m+n)^2$, since the amplifier is applied to the 10\,GHz signal prior to EO multiplication. Thus, full cancellation of \( S_{\phi,\mathrm{HPA}}(f) \) requires that the DRO signal be tapped at a point that includes all noise sources present in the modulation and feedback loop~\cite{li2014electro}. This condition is easily satisfied in architectures using a single HPA.

Therefore, moderate optical spans (e.g., 300\,GHz) with low \( S_{\phi,\Delta \nu}(f) \) offer a key architectural advantage: they enable the use of a single EOM and HPA, allowing complete rejection of amplifier phase noise and optimizing spectral purity in the divided microwave signal. 

\section{Electro-Optic Division of a Dual-Wavelength Brillouin Laser}

\subsection{Setup}

\begin{figure}[htbp]
    \centering
    \includegraphics[width=\textwidth]{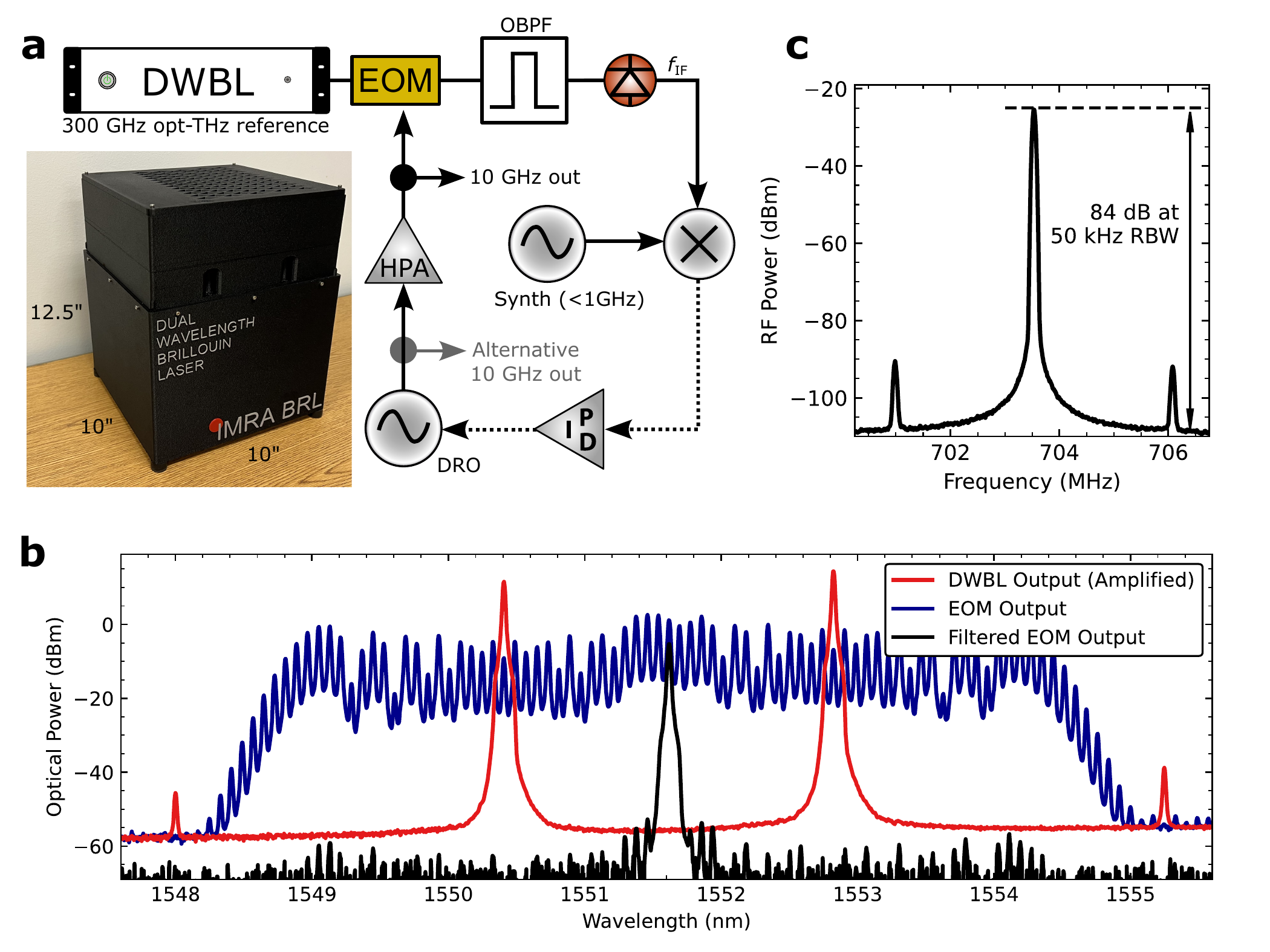}
    \caption{Setup used for eOFD using a DWBL. \textbf{(a)} Schematic of the full eOFD system and photo of the DWBL used. \textbf{(b)} Optical spectra: optically amplified DWBL output (red), EO-modulated output (blue), and filtered sidebands (black). \textbf{(c)} RF spectrum of the detected EO comb beatnote (RBW: 50\,kHz).}
    \label{fig:SchematicEOFD}
\end{figure}

We implement eOFD using a dual-wavelength Brillouin laser (DWBL) operating at 300\,GHz~\cite{heffernan_brillouin_2024} to synthesize a 10\,GHz signal with ultra-low phase noise. The schematic in Fig.~\ref{fig:SchematicEOFD}(a) shows the DWBL (pictured) whose output passes through an EOM driven by a low-noise, amplified DRO. High-order optical sidebands are generated, and an optical bandpass filter (OBPF) isolates a target sideband pair. Photodetection of these filtered tones yields an intermediate frequency \( f_{\mathrm{IF}} \), which is mixed with a reference signal \( f_{\mathrm{ref}} \) to produce an error signal for phase-locking the dielectric resonator oscillator to the DWBL.

A key design detail is the placement of the 10\,GHz output tap \textit{after} the HPA, which minimizes added phase noise from the RF drive chain. The DWBL used here (20\,L volume including power supplies and real-time monitoring electronics) also represents a fivefold size reduction compared to the previous 350\,L system~\cite{heffernan_brillouin_2024}. Engineering the thermal management of the spool has eliminated the need for the rough vacuum chamber previously required for precise temperature control. Additionally, the adoption of vibration-canceling spool, following the design in Ref.~\cite{jeon_palm-sized_2023}, has removed the need for a vibration isolation table, dramatically reducing SWaP and system complexity.

Figure~\ref{fig:SchematicEOFD}(b) highlights the effectiveness of using a single EO modulator to generate a 300\,GHz optical span. Cascading multiple modulators introduces additional insertion loss; therefore, if the modulator's $V_{\pi}$ is sufficiently low, generating the full span with a single device can preserve optical SNR for detection. The red trace shows the amplified DWBL output, the blue trace the modulated signal, and the black trace the resulting EO combs from both optical lines exhibit excellent overlap, yielding a high optical SNR of approximately 60\,dB.

The corresponding intermediate frequency (IF) beatnote, shown in Fig.~\ref{fig:SchematicEOFD}{c}, exhibits an SNR of 84\,dB within a 50\,kHz resolution bandwidth. Such high SNR is crucial to minimizing in-loop error in the PLL, particularly since the division ratio from 300\,GHz to 10\,GHz offers only limited suppression of residual noise. During operation, the DRO is locked using a 704\,MHz reference (Keysight MXG) and a 2\,MHz-bandwidth PLL (IMRA ULC), with control applied directly to the tuning port of the DRO (InWave DRO-1019).

\subsection{Phase Noise Performance}

\begin{figure}[htbp]
    \centering
    \includegraphics[width=\figwidthPN]{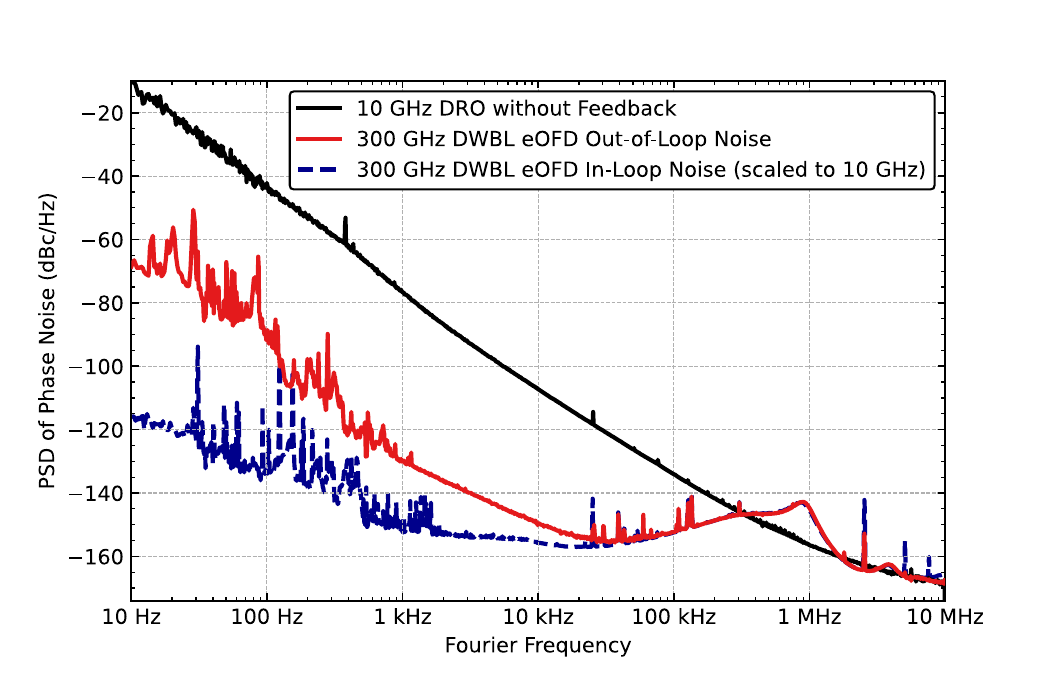}
    \caption{One-sided phase noise PSD of the 10\,GHz output derived from the DWBL using eOFD. The native DRO signal (black), out-of-loop locked output (red), and in-loop error signal (blue) are shown.}
    \label{fig:10Ghz}
\end{figure}

Figure~\ref{fig:10Ghz} presents the phase noise PSD of the 10\,GHz signal synthesized from the 300\,GHz dual-wavelength Brillouin laser using the eOFD technique. Phase noise measurements in this work were performed using the Keysight SSA-X, which implements cross-correlation techniques with two independent local oscillators. The phase noise of the free-running DRO (black) is shown alongside the out-of-loop phase noise of the eOFD-stabilized output (red), and the in-loop error signal (blue dashed), numerically scaled from 300\,GHz down to the 10\,GHz carrier for direct comparison. The locked signal exhibits excellent spectral purity across the full Fourier frequency range, demonstrating the efficacy of the eOFD architecture in preserving the low-noise properties of the optical reference. Specifically, the measured phase noise reaches --130\,dBc/Hz at a 1\,kHz offset, --150\,dBc/Hz at 10\,kHz, and --170\,dBc/Hz at 10\,MHz. These values represent a substantial improvement over the unlocked DRO within the loop bandwidth and confirm that the division process introduces minimal additional noise when properly implemented. 

While the servo bump remains below -140\,dBc/Hz, further reduction could unlock even cleaner mid-band (100 kHz - 1 MHz) performance. Several strategies could be employed to reduce it. First, increasing the SNR of the intermediate frequency \( f_{\mathrm{IF}} \) would allow for a wider loop bandwidth, enabling better correction of DRO noise at higher Fourier frequencies. In turn, this would result in a reduced magnitude of the servo bump, because it would appear on the DRO at a higher Fourier frequency, where the phase noise is lower to begin with. A DRO with improved free-running phase noise, if commercially available, or a higher frequency division ratio would also mitigate this issue. Another potential solution involves applying the in-loop error signal (i.e., the DC output of the mixer) directly to the DRO through a phase shifter in a feedforward configuration~\cite{zhang2011advanced}. While optimization of these approaches was not the focus of this work, we note that viable pathways exist to improve this aspect of the system performance.

\subsection{Noise Source Analysis in eOFD-Stabilized Systems}

\begin{figure}[htbp]
    \centering
    \includegraphics[width=\figwidthPN]{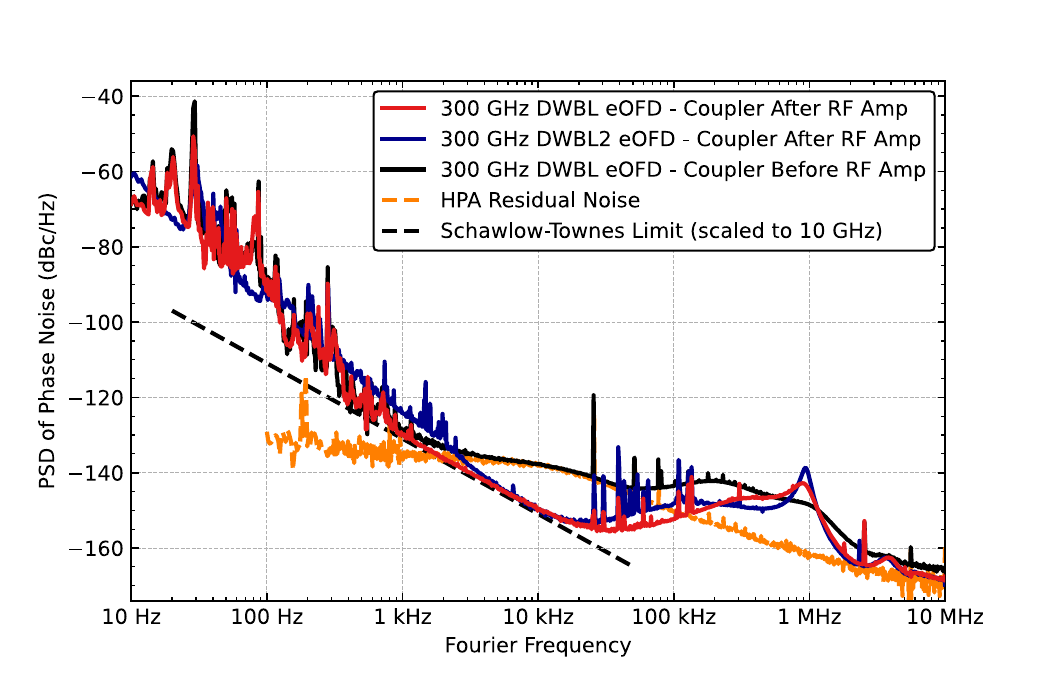}
    \caption{Breakdown of phase noise contributions in the eOFD-stabilized DRO system. Comparison of different DWBL configurations, output tap points, and noise mechanisms: DWBL noise with a tap-off after the HPA (red), before the HPA (black), after the HPA for a larger DWBL design presented previously (blue)~\cite{heffernan_brillouin_2024}, theoretical noise floor set by Brillouin cavity (black dashed), and the residual noise of the HPA (orange). DWBL noise is dominated by acoustic coupling at low frequency, the servo bump near the loop bandwidth, and intrinsic DRO noise at high offsets.}
    \label{fig:10GHzNoise}
\end{figure}

Phase noise measurements across different configurations and DWBL systems are shown in Fig.~\ref{fig:10GHzNoise}. The minor design updates in the 20\,L DWBL (red) led to slightly lower noise below 3\,kHz compared to the "DWBL2" system (blue) used in Ref.~\cite{heffernan_brillouin_2024}. The black trace corresponds to a configuration where the 10\,GHz signal is extracted \textit{before} the HPA. In this setup, amplifier noise is not suppressed by the loop and appears as a significant elevation in phase noise between 2\,kHz and 100\,kHz, when compared against optimal tap placement for suppressing phase noise (red and blue traces). For confirmation, the residual phase noise of the HPA alone is plotted in orange and matches the observed degradation in the black trace. Discrepancies between the pre-HPA and post-HPA traces at high Fourier frequency ($> 100$ kHz) are due to the different PID settings used between the measurements.

The optimal tap configuration for reducing phase noise involves splitting the \textit{amplified} 10\,GHz DRO output \textit{after} the HPA. This ensures that the amplifier lies within the feedback loop, allowing its noise to be corrected by the PLL. The result is a significantly cleaner phase spectrum, with noise levels approaching the fundamental limit set by the Brillouin laser cavity, namely the Schawlow-Townes limit (black dashed line).

The Schawlow–Townes limit for Brillouin lasers differs from the limit of most other lasers, as spontaneous emission is not the dominant noise mechanism. Instead, the fundamental phase noise floor is governed by thermally populated phonon modes, which correspond to incoherent perturbations in the density wave that mediates the Brillouin gain. The one-sided phase noise PSD from the Schawlow-Townes effect in a Brillouin laser near room temperature is given by \cite{Li2013BrillouinChip},

\begin{equation} \label{eq:ST}
    S_\phi^\mathrm{ST} (f) \approx \frac{c k_\mathrm{B} T}{8 \pi^2 \tau \tau_\mathrm{ex}P n v_\mathrm{a}} f^{-2} ,
\end{equation}

\noindent where $c$ is the speed of light in vacuum, $k_\mathrm{B}$ is the Boltzmann constant, $T$ is the temperature, $\tau$ is the photon lifetime in the fiber cavity, $\tau_\mathrm{ex}$ is the lifetime neglecting internal losses, $P$ is the output power of the laser, $n$ is the index of refraction, and $v_a$ is the acoustic velocity. Note that Eq. (\ref{eq:ST}) is for a single-wavelength Brillouin laser, and a DWBL exhibits a phase noise a factor of 2 higher, assuming identical cavity parameters at the two wavelengths. Additional contributions arise from technical noise sources such as environmentally induced fiber length variations and phase noise transferred from the pump \cite{Debut2000}. These effects determine the fundamental noise floor of the DWBL system. Schawlow–Townes noise from the two optical tones is not suppressed through common mode rejection, since each Brillouin line is generated independently through separate Stokes processes. As a result, their beatnote preserves uncorrelated noise contributions, including residual phonon and pump-induced fluctuations. In the present system, the phase noise floor, scaled to 10\,GHz, is measured at $\left(-70 - 20 \log_{10} (f/1 \, \mathrm{Hz}) \right) \,\mathrm{dBc/Hz}$ in excellent agreement with Eq. \ref{eq:ST} (dashed black line on Fig.~\ref{fig:10GHzNoise}).

\subsection{Architectural Tradeoffs}

\begin{figure}[htbp]
    \centering
    \includegraphics[width=\figwidthPN]{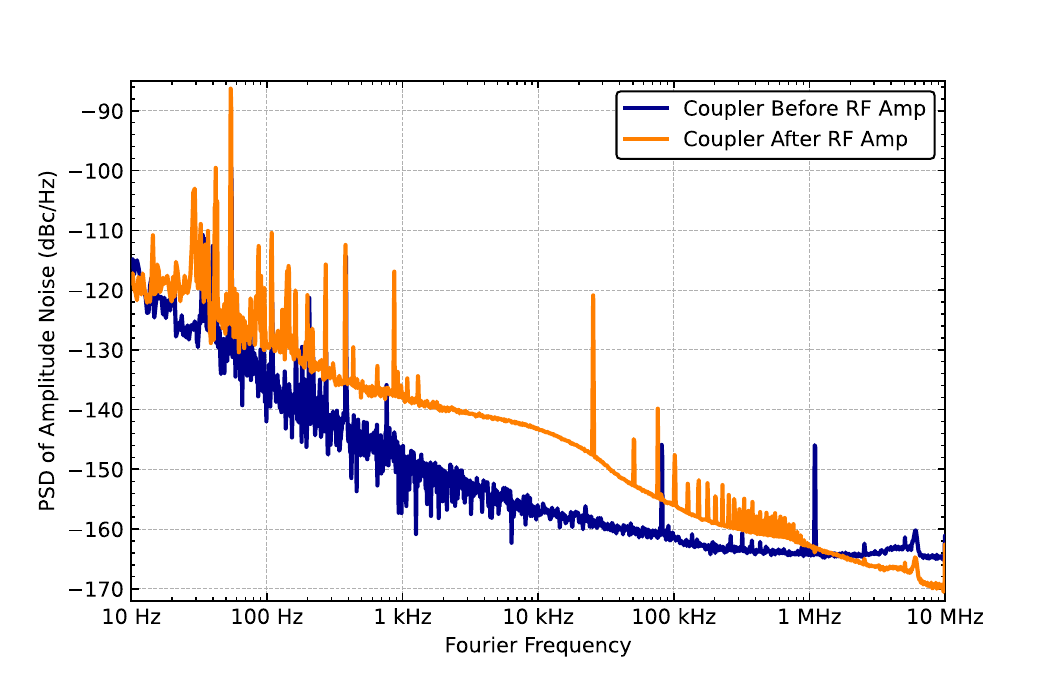}
    \caption{One-sided amplitude noise PSD of the 10\,GHz output of the DRO as measured before (blue) and after (orange) the RF HPA}
    \label{fig:AN}
\end{figure}

While placing the RF output coupler after the HPA enables effective suppression of amplifier-induced phase noise, this configuration inevitably introduces a tradeoff with respect to amplitude noise that has not previously been explored when discussing HPA placement in eOFD systems~\cite{li2014electro}. A PLL detects and corrects phase fluctuations between a signal and a reference and is completely insensitive to amplitude variations. As such, amplitude noise - originating from sources such as thermal fluctuations, nonlinearities, or power supply ripple within the HPA - cannot be corrected by the PLL and is instead directly transferred to the output. In this configuration, any residual amplitude noise added by the HPA is therefore imprinted on the final 10\,GHz signal. As shown in Fig.~\ref{fig:AN}, this results in an \emph{excess} amplitude noise of approximately 14\,dB at a 10\,kHz Fourier offset (orange) when compared to the pre-HPA configuration (blue). 

This tradeoff highlights an important design consideration: while placing the output tap after the HPA yields optimal suppression of phase noise, crucial for applications such as coherent communication and time-frequency metrology, it comes at the cost of increased amplitude noise. In systems where amplitude noise is equally critical, such as radar or analog photonic links, this tradeoff becomes nontrivial. Since a single tap point cannot simultaneously optimize both, designers must deliberately prioritize which noise contribution has the greater impact on overall system performance.

While the ability to perform eOFD using only a 300\,GHz optical span represents a major architectural simplification, it is important to emphasize that the DWBL is not fundamentally limited to this frequency. In fact, the same DWBL architecture supports optical spacings into the multi-terahertz regime. To explore this potential, we characterized the phase noise of a 3.33\,THz DWBL using a fiber-delay-line-free optical-RF heterodyne technique. This enables us to assess its suitability as a reference for high-division-ratio eOFD, while also motivating the exploration of alternatives to conventional EO combs capable of efficiently bridging multi-THz spans while maintaining sufficient SNR.

\section{Optical-RF Heterodyne Technique for Phase Noise Measurement without Fiber Delay Lines}

\subsection{Derivation}

To project 10\,GHz phase noise performance of a 3.33\,THz DWBL eOFD system, it is necessary to first characterize its phase noise at the native optical frequency spacing. However, direct phase noise measurements at multi-THz frequencies are not feasible due to the limitations of available photodetectors and mixers. Additionally, indirect techniques based on two-wavelength fiber delay lines suffer from thermomechanical fiber noise, especially for long delays needed to access low Fourier frequencies~\cite{kuse2018photonic}. This severely limits their ability to resolve intrinsic noise at sub-kHz offsets.

To overcome this, we developed an optical-RF heterodyne approach inspired by the two-wavelength differential interferometry (TWDI) concept but implemented without any fiber delay. The key idea is to compare two DWBLs with aligned optical frequencies and nearly identical opto-THz spacing. Since both DWBL emit a pair of optical tones near 1550\,nm, their beatnote differences carry information about the relative phase noise of their THz spacing without requiring direct THz detection.

The optical-RF heterodyne setup is illustrated optically in Fig.~\ref{fig:Heterodyne}(a), with an overview of RF components in (b). Referring back to Eq. (\ref{eq:nu}), let the first DWBL emit two optical tones \( \nu_1 \) and \( \nu_2 \), and the second DWBL emit \( \nu_1' \) and \( \nu_2' \). The two DWBL are combined on a single photodetector, producing two intermediate-frequency beatnotes,
\begin{align}
f_{\mathrm{IF,1}} &= \nu_1 - \nu_1', \\
f_{\mathrm{IF,2}} &= \nu_2 - \nu_2',
\end{align}

\noindent in the RF domain (hundreds of MHz to a few GHz) such that they can be detected with standard photodiodes. They are then electrically mixed to yield,
\begin{equation}
f_{\mathrm{mix}} = (\nu_1 - \nu_1') - (\nu_2 - \nu_2') = (\nu_1 - \nu_2) - (\nu_1' - \nu_2').
\end{equation}

This output corresponds to the difference between the THz beatnotes of the two DWBLs. Let \( \phi_1(t), \phi_2(t) \) and \( \phi_1'(t), \phi_2'(t) \) be the instantaneous phases of the four optical tones. The phase fluctuation of the mixed signal is then given by,
\begin{equation}
\phi_{\mathrm{mix}}(t) = \left[ \phi_1(t) - \phi_2(t) \right] - \left[ \phi_1'(t) - \phi_2'(t) \right]
= \phi_{\Delta \nu}(t) - \phi_{\Delta \nu'}(t),
\end{equation}

\noindent with \( \phi_{\Delta \nu}(t) \) and \( \phi_{\Delta \nu'}(t) \) representing the phase fluctuations signals for each DWBL, respectively. Therefore, the corresponding phase noise PSD is proportional to,
\begin{equation}
S_{\phi,\mathrm{mix}}(f) \propto S_{\phi,\Delta \nu}(f) + S_{\phi,\Delta \nu'}(f),
\end{equation}

\noindent where \( S_{\phi,\Delta \nu}(f) \) and \( S_{\phi,\Delta \nu'}(f) \) are the phase noise PSDs of the THz spacing from each DWBL, assuming no correlation between \( S_{\phi,\Delta \nu}(f) \) and \( S_{\phi,\Delta \nu'}(f) \). Thus, the optical-RF heterodyne method yields the sum of the phase noise PSDs from both DWBLs, or nominally 3\,dB higher than a single DWBL if performance is comparable. 

This technique provides a reliable way to characterize opto-terahertz sources at very high frequency spacing using only standard photodiodes and RF electronics, and without fiber-delay-line-induced limitations. By comparing directly with millimeter-wave measurements at 300\,GHz, we validate our new technique for measuring the projected 3.33\,THz phase noise, enabling extrapolation to future microwave division results such as 10\,GHz eOFD of a 3.33\,THz DWBL.

\subsection{Phase Noise Measurements}

\begin{figure}[htbp]
\centering
\includegraphics[width=\figwidthPN]{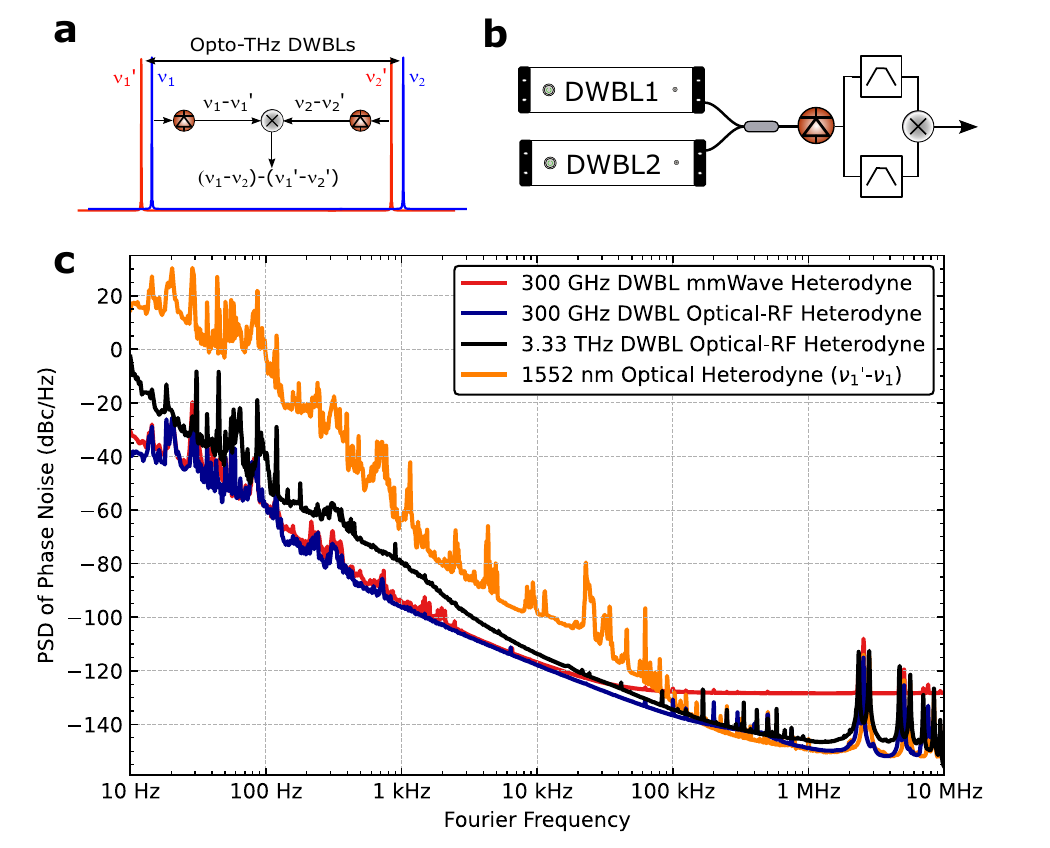}
\caption{Comparison of heterodyne DWBL characterization techniques. (\textbf{a}) optical and (\textbf{b}) RF signals diagrams of the optical-heterodyne DWBL measurement technique. Two DWBLs at similar opto-THz and optical frequencies are mixed onto a single photodiode, generating heterodyne beat signals at the differences between each pair of corresponding optical frequencies. These beat notes are then heterodyned to generate the difference frequency between the two opto-THz DWBL sources. (\textbf{c}) One-sided PSD of phase noise for the two DWBLs at 300\,GHz measured directly using millimeter-wave components (red) and the new optical-RF heterodyne (blue) techniques, highlighting the comparable results out to the white noise floor of the 300\,GHz fundamental mixer (\(-128\,\mathrm{dBc/Hz}\)). Phase noise for two DWBL at 3.33\,THz is shown in black. A direct optical beatnote, $\nu_1-\nu_1'$, emphasizes the 50\,dB noise rejection of the dual-heterodyne technique (orange).
\label{fig:Heterodyne}
}
\end{figure}

Figure~\ref{fig:Heterodyne}(c) shows the one-sided phase noise PSD obtained using the optical-RF heterodyne technique. To validate this method, we compared it against a direct millimeter-wave measurement performed on a pair of 300\,GHz DWBLs using uni-traveling-carrier photodiodes (UTC-PDs) followed by waveguide-based millimeter-wave mixing~\cite{heffernan_brillouin_2024}. The blue trace corresponds to the optical-RF heterodyne measurement, while the red trace represents the direct measurement using the traditional UTC-PD and mm-wave mixer setup. 

The two results agree across the entire 10\,Hz–10\,kHz Fourier frequency range, confirming the validity of the optical-RF heterodyne technique. At Fourier frequencies above 20\,kHz, discrepancies between the two methods are dominated by the detection noise floor (\(-128\,\mathrm{dBc/Hz}\)), primarily due to the high conversion loss of the fundamental millimeter-wave mixer.

At low offset frequencies ($\lesssim 3$kHz), both traces exhibit noise peaks we attribute to fiber cavity length variations caused by the acoustic and thermal environments. At higher frequencies, the optical-RF method is 20\,dB below the noise floor of the mm-wave measurement (\(-128\,\mathrm{dBc/Hz}\)), highlighting its superior sensitivity and utility for characterizing terahertz phase noise without requiring broadband detection.

The method is then extended to a pair of 3.33\,THz DWBL systems, shown in black in Fig.~\ref{fig:Heterodyne}(c). Since terahertz detectors and components exhibit vanishing sensitivity at an IF of 3.33\,THz, this optical-RF heterodyne technique uniquely enables low-phase-noise characterization in this frequency regime. The black trace shows elevated phase noise across all offset frequencies, consistent with the expected scaling: common-mode rejection of fiber cavity length fluctuations is a factor of $(3.33/0.3)^2$ less effective, as compared to the 300\,GHz case. Despite this increase, the characteristic spectral shape remains unchanged, exhibiting a vibration-dominated regime at low Fourier frequencies and a flat noise floor beyond 1\,MHz. The phase noise approaches --150\,dBc/Hz at 1\,MHz offset, representing a record performance at multi-terahertz frequencies. Strong side-mode oscillations at the fiber cavity’s free spectral range are visible in the high-frequency region and ultimately limit the achievable phase noise beyond 1\,MHz. Due to differing free spectral ranges for DWBL1 and DWBL2 used in these comparative measurements, there are two sets of side-mode oscillations.

We also include the PSD of the direct optical beatnote, $(\nu_1'-\nu_1)$, in orange, to emphasize the effectiveness of the optical-RF heterodyne technique in rejecting common-mode phase noise between the two Brillouin lines.

This optical-RF heterodyne technique thus provides a robust, fiber-delay-line-free, and scalable tool to evaluate the phase noise of opto-THz oscillators across THz spacings. This makes it, to our knowledge, the only delay-line-free technique for direct relative phase noise evaluation in the multi-terahertz domain. It also offers a critical path forward for assessing the feasibility of electro-optic frequency division (eOFD) from 3.33\,THz down to 10\,GHz. Generating an EO comb that spans this bandwidth directly is not currently practical using a single bulk EOM, as the required modulation index would necessitate cascading multiple stages with multiple uncorrelated HPAs, resulting in uncompensated phase noise and high optical insertion loss.

Nonetheless, eOFD using larger division factors remains enticing, especially when leveraging the demonstrated stability of the 3.33\,THz spacing shown in Fig.~\ref{fig:Heterodyne}(c). Recent developments in low-\(V_\pi\) modulators~\cite{cheng2025spiral} and Kerr injection in integrated microresonator combs~\cite{sun2024kerr} offer the potential to bridge large optical gaps with reduced complexity and footprint, enabling high-division-factor eOFD architectures that maintain exceptional spectral purity without relying on broadband photodetection or millimeter-wave synthesis chains.

\section{Comparison of 10\,GHz Phase Noise}

\begin{figure}[htbp]
\centering
\includegraphics[width=\figwidthPN]{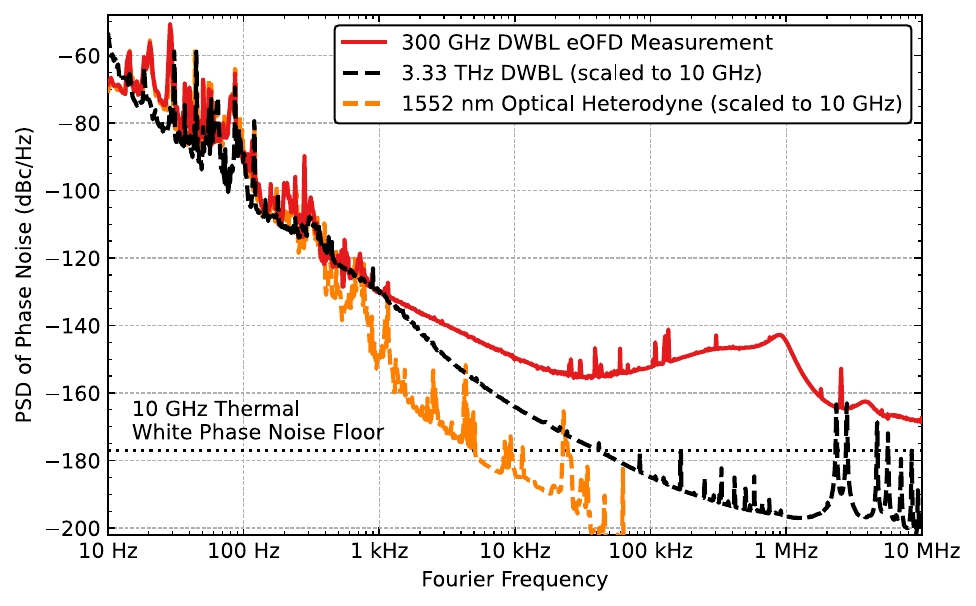}
\caption{Comparison of three OFD techniques for generating a 10\,GHz source. 
300\,GHz DWBL (red) eOFD as measured in this work, 
3.33\,THz DWBL (black dashed) scaled to 10\,GHz using a yet-to-be-implemented division scheme, 
1552\,nm Optical Heterodyne (orange dashed) of a single Brillouin laser scaled to 10\,GHz, for example using OFD of a self-referenced frequency comb.  
The impact of acoustic noise ($f$<10\,kHz) highlights the limit of the optical cavity independent of OFD technique. The thermal white phase noise floor (--177 dBc) is also shown for reference as a lower bound on achievable high-frequency performance. 
\label{fig:Comparison}
}
\end{figure}

To contextualize the performance of this 300\,GHz DWBL-based eOFD architecture, in Fig.~\ref{fig:Comparison} we compare the measured 10\,GHz phase noise (red) against hypothetical, ideal frequency division of the 3.33\,THz DWBL (black dashed) and 1552\,nm optical heterodyne of a single Brillouin source (orange dashed) measurements that were shown in Fig.~\ref{fig:Heterodyne}(c). 

 Since eOFD of the 3.33\,THz would suppress all phase noise contributions (except that of the HPA) by a factor of $(m+n)^2$, increasing optical span by a factor of eleven would enable significantly lower phase noise across the entire offset range above 10\,kHz. Beyond 100\,kHz, The projected 3.33\,THz trace dips below even the thermal white phase noise floor for a CW 10\,GHz oscillator. As highlighted previously, achieving this performance remains contingent on overcoming the considerable technical challenges of generating and detecting a high-SNR beatnote across a 3.33\,THz span.

As a further perspective, we include in orange the optical phase noise of the 1552\,nm Brillouin lines, scaled down to 10\,GHz (achievable by means of, for instance, a self-referenced optical frequency comb). We note that, due to common mode rejection, the phase noise for the 300\,GHz, 3.33\,THz, and optical references have essentially identical performance below 500\,Hz. This suggests that, for applications dependent on low-frequency performance, there is no advantage to be had by exploiting an optical rather than an opto-terahertz reference, despite the much greater complexity the former demands.

This comparison illustrates the compelling performance-to-complexity tradeoff achieved by this 300\,GHz DWBL eOFD system and future, THz-spanning systems. It combines near state-of-the-art phase noise with a streamlined architecture, offering a scalable path toward compact and high-purity microwave generation.

\section{Conclusion}

We have demonstrated a compact, high-performance eOFD architecture that synthesizes a 10\,GHz microwave signal from a 300\,GHz DWBL. By leveraging the exceptional spectral purity of the DWBL and a minimalist configuration using a single EO phase modulator, the system achieves phase noise levels of --130\,dBc/Hz at 1\,kHz, --150\,dBc/Hz at 10\,kHz, and --170\,dBc/Hz at 10\,MHz. These results match or exceed the performance of traditional OFD systems while significantly reducing complexity, component count, and footprint.

This architecture eliminates the need for self-referenced combs, cascaded modulators, and nonlinear fiber, marking a major simplification in low-noise microwave synthesis and offering a scalable path toward higher division factors. By extracting the output 10\,GHz signal after the HPA, amplifier phase noise is fully suppressed, preserving the optical coherence of the DWBL at microwave frequencies. The impact of this decision on amplitude noise was also explored. 

We have further shown that extending the optical span to 3.33\,THz holds the potential for record-breaking phase noise performance. While current limitations in electro-optic comb bandwidth and beatnote detection present technical challenges, the demonstrated phase stability points to a clear roadmap. The integration of low-$V_\pi$ modulators, high-power RF drivers, and advanced photonic packaging will be key to realizing next-generation eOFD systems with high division ratios and minimal SWaP requirements.

Although phase noise has been the primary focus of most eOFD demonstrations, long-term frequency stability remains equally critical for many applications. In this work, the stability of the synthesized 10\,GHz signal is directly inherited from the opto-terahertz reference, characterized in Ref.~\cite{heffernan_brillouin_2024}. To further enhance long-term stability, one promising approach is to frequency-lock the opto-terahertz beatnote to a molecular rotational transition in the terahertz range. As demonstrated in Refs.~\cite{greenberg2024terahertz,greenberg2025dual}, locking an opto-THz oscillator to a molecular line - such as those in N$_2$O or OCS - can yield fractional frequency stabilities on the order of $1 \times 10^{-12}$, and potentially orders of magnitude lower~\cite{mcgrew2025terahertz}, at one second of averaging. When combined with eOFD, this would enable a 10\,GHz output with both ultra-low phase noise and exceptional long-term stability. Such a system would transition DWBL-referenced oscillators from transfer tools into fully-referenced microwave secondary frequency standards.

This work establishes the dual-wavelength Brillouin laser as a transformative opto-terahertz reference and positions electro-optic frequency division as a powerful, practical approach to generating low-noise microwave signals. The demonstrated architecture opens the door to deployable, compact systems for timing, communication, and spectroscopy, with levels of performance previously limited to laboratory-scale instruments.

\section*{Acknowledgments}

We thank John Dorighi at Keysight Technologies for generously providing access to the SSA-X cross-correlator-based phase noise analyzer which enabled several key measurements in this work. We also acknowledge Hideyuki Ohtake and Yuki Ichikawa for their support and helpful discussions throughout the project.

\bibliography{references}

\end{document}